\documentstyle[prb,aps,epsfig]{revtex}

\begin{document}
\title{Impurity state in the vortex core of $d$-wave superconductors:
Anderson impurity model versus unitary impurity model}
\author{Qiang Han,$^{1,2}$ Z. D. Wang,$^{1,2,3}$\thanks{To whom correspondence should be addressed.
Email address: zwang@hkucc.hku.hk} X. -G. Li,$^{1}$, and Li-yuan
Zhang$^{4}$}
\address{$^{1}$Structure Research Laboratory, Department of Material Science and Engineering, University of
Science and Technology of China, Hefei Anhui 230026, China}
\address{$^{2}$Department of Physics, University of Hong Kong, Pokfulam Road, Hong Kong, China}
\address{$^{3}$Texas Center for Superconductivity, University of Houston, Texas 77204, USA}
\address{$^{4}$Department of Physics, Peking University, Beijing 100871, China}
\maketitle

\begin{abstract}
Using an extended Anderson/Kondo
 impurity model to describe the magnetic moments  around an impurity doped in
high-$T_{\text{c}}$ $d$-wave cuprates and in the framework of the slave-boson
meanfield approach, we study numerically the impurity state in the vortex core
by exact diagonalization of the well-established Bogoliubov-de Gennes equations.
The  low-energy impurity state is found to be
in good agreement with scanning tunneling
microscopy observation. After pinning a vortex on the impurity
site, we compare the  unitary impurity model with
the extended Anderson impurity model by examining the effect
of the  magnetic field on
the impurity state. We
find that the impurity resonance in the unitary impurity model is
strongly suppressed by the vortex; while it is insensitive to the
field in the extended Anderson impurity model.
\end{abstract}

Effects of nonmagnetic impurities, such as Zinc, doped in the
CuO$_2$ plane of high-$T_{\text{c}}$ materials have attracted much
attention in recent years as it may help us
to better understand the underlying mechanism of
high-$T_{\text{c}}$ superconductivity
\cite{balatsky-nature-403-717-2000}. The low-energy (near zero)
impurity resonant state in $d$-wave superconductors was first
predicted in Refs.
\cite{balatsky-prb-51-15547-1995,salkola-prl-77-1841-1996,flatte-prb-56-11213-1997}
based on a unitary scattering-potential impurity model and later
successfully observed in a series of beautiful atomic-scale
scanning tunneling microscopy (STM) experiments
\cite{hudson-sci-285-88-1999,yazdani-prl-83-176-1999,pan-nature-403-746-2000}
in the vicinity of individual Zn ions in
Bi$_2$Sr$_2$Ca(Cu$_{1-x}$Zn$_x$)$_2$O$_{8+\delta}$. Since
Zn$^{++}$ has no spin itself (therefore nonmagnetic), it is
natural to treat the impurity as a point-like scalar potential scatterer of
conduction electrons. Indeed,  both energy position and four-fold
symmetric spatial distribution \cite{pan-nature-403-746-2000} of
local density of states (LDOS) of quasiparticle can be explained
consistently by theoretical calculations based on the $t$-matrix
scattering potential theory  in the unitary limit
\cite{salkola-prl-77-1841-1996}. Nevertheless, the sign and magnitude of
the scattering potential as well as the particle-hole symmetry play a
crucial role in determining the exact energy level of impurity
resonant state relative to Fermi energy. The scattering potential
chosen in various theoretical studies differ largely from $-4.4$
eV to $18.9$ eV
\cite{zhu-prb-64-060501-2001,polkovnikov-prl-86-296-2000,martin-prl-88-097003-2002}.
The continuum $t$-matrix theory, which assumed particle-hole
symmetry and {\it repulsive} potential, predicted impurity state
with energy \cite{salkola-prl-77-1841-1996} consistent with STM
observation \cite{pan-nature-403-746-2000} that the resonance
lies slightly below the Fermi level; on the other hand, the
particle-hole symmetry is not applicable to high-$T_c$ cuprates
 and
theoretical calculations
\cite{polkovnikov-prl-86-296-2000,martin-prl-88-097003-2002}
show that in the absence of such symmetry  only strong
{\it attractive} potential can produce impurity state with
negative energy while repulsive potential
\cite{martin-prl-88-097003-2002} gives positive energy contrary
to experiment results. Furthermore, the unitary impurity model
has difficulty in accounting for spectra distribution pattern of
differential conductance at the resonance energy. The experiment shows
that the zero bias peak is strongest on the Zn ion  with local
maximum  on its second nearest neighbor Cu sites
\cite{pan-nature-403-746-2000} while the unitary impurity model
concluded that the spectral weight on Zn atom is negligibly small
and local maximum
peaks are on nearest neighbor Cu sites of the impurity
\cite{salkola-prl-77-1841-1996,zhu-prb-62-6027-2000,martin-prl-88-097003-2002}.
In order to reconcile the apparent discrepancy, the effect of the
interlayer tunneling matrix elements, such as a blocking effect of
BiO layer \cite{zhu-prb-62-6027-2000} and a fork model
\cite{martin-prl-88-097003-2002} have been employed to reproduce
the distribution of spectral weight .
At present, whether the spatial pattern of the tunneling
conductance of the impurity resonance is intrinsic in the CuO$_2$
plane or interlayer tunneling effect must be considered is still
an open question for both theoretical investigation and further
experimental examination. Recently, several alternative
theoretical models
\cite{zhu-prb-64-060501-2001,polkovnikov-prl-86-296-2000,nagaosa-prl-79-3755-1997,zhu-prb-63-020506-2000}
have been suggested,
 which take the effect of the Kondo screening
of the local magnetic moments induced around the Zinc atom into
account, mostly motivated by the observation of nuclear magnetic
resonance (NMR)
\cite{bobroff-prl-83-4381-1999,julien-prl-84-3422-2000,bobroff-prl-86-4116-2001}
and neutron scattering experiments
\cite{fong-prl-82-1939-1999,sidis-prl-84-5900-2000} that
staggered magnetic moments are developed around Zn ion with total
net spin $1/2$. When Kondo spin dynamics of such moment is
considered, by including an exchange interaction between the
induced spin and the spin of conduction electrons in the model
Hamiltonian, Polkovnikov {\it et al}
\cite{polkovnikov-prl-86-296-2000} produce an impurity state
whose energy dependence and spatial pattern may fit well with STM
spectra, without a  strong scattering potential
and the specific characteristic of interlayer tunneling matrix
elements.

Motivated by the above controversy between the unitary impurity
model and the model invoking induced magnetic moments, in this
work we intend to compare these two models in the case that a
single vortex line is pinned by the impurity. To describe the
Kondo effect of the magnetic moments and the effect of vortex,
 we introduce an extended Anderson impurity model
with a phase factor being dependent on site and vortex.
 In the absence of vortex, both models with appropriate parameters can generate
low-lying impurity state. However, when a vortex is present, for
the unitary impurity model, the pronounced LDOS peak generated
around the Zn impurity is largely decreased or even destroyed,
indicating the vanishing of the impurity resonance by the vortex
core; while for the extended Anderson impurity model, the
existence of the vortex center has a weak effect on the impurity
state. Such a difference may be examined
readily by STM experiments on
Bi$_2$Sr$_2$Ca(Cu$_{1-x}$Zn$_x$)$_2$O$_{8+\delta}$ samples subject to
a strong magnetic field.


In this work, we adopt an extended Hubbard model on a two
dimensional (2D) lattice with nearest-neighbor (NN) hopping and NN
pairing interaction to model the $d$-wave high-$T_{\text{c}}$
cuprates and the extended Anderson impurity model to describe the
magnetic moments  around the impurity(Zinc ion) doped in the
CuO$_2$ plane. The model Hamiltonian is expressed as
\begin{equation}
H=H_{\text{dsc}}+H_{\text{imp}}
\end{equation}
where
\begin{equation}
H_{\text{dsc}}=-\sum_{\langle i,j\rangle
\sigma}t_{ij}c_{i\sigma}^{\dagger}c_{j\sigma}+ \sum_{\langle
i,j\rangle
\sigma}(\Delta_{ij}c_{i\uparrow}^{\dagger}c_{j\downarrow}^{\dagger}+\text{h.c.})-
\sum_{i,\sigma}\mu c_{i\sigma}^{\dagger}c_{i\sigma},
\end{equation}
and
\begin{equation}
H_{\text{imp}}=H_{\text{mag}}+
V_I
c_{0\sigma}^{\dagger}c_{0\sigma}. %
\label{himpurity}
\end{equation}
Here $H_{\text{dsc}}$ is the BCS-like Hamiltonian of the host
$d$-wave superconductor described on a 2D lattice. $\langle
i,j\rangle $ refers to the NN sites and $t_{ij}$ the hopping
integral between site $i$ and $j$. $\mu$ is the chemical
potential. $\Delta_{i,j}$ is the bond paring potential defined as
$\Delta_{ij}=-V_d\langle c_{i\downarrow}c_{j\uparrow}\rangle $
with $V_d$ the effective pairing strength between electrons on NN
sites. $H_{\text{imp}}$ represents the impurity Hamiltonian
which includes   both an on-site scattering potential term
represented by $V_I$ and the term  $H_{\text{mag}}$
describing the local moments around the impurity.
  As suggested by NMR experiments
\cite{bobroff-prl-83-4381-1999,julien-prl-84-3422-2000,bobroff-prl-86-4116-2001},
the induced magnetic moments with net spin $1/2$ are mainly
located at the four NN Cu sites of the central Zn impurity site
${\bf r}_0=(0,0)$ while negligible right on the Zn site; therefore
as in Ref. \cite{polkovnikov-prl-86-296-2000}, we assume that the
magnetic impurity with
 an effective $1/2$ spin may reside only in the NN sites  and couples with
conduction electrons on the NN sites. Namely, the $H_{\text{mag}}$
may be modeled by the extended Anderson impurity model with strong
Hubbard repulsion $U_d$ as $ H_{\text{mag}}=\sum_{\sigma}
\epsilon_d d_{\sigma}^{\dagger}d_{\sigma}+ U_d
d_{\uparrow}^{\dagger}d_{\uparrow}d_{\downarrow}^{\dagger}d_{\downarrow}+
\sum_{{\bf r}_i={\boldmath{\mbox{$\tau$}}},\sigma}
V_h(i)(e^{\phi_i} c_{i\sigma}^{\dagger}d_{\sigma} + \text{h.c.}) $
with ${\boldmath{\mbox{$\tau$}}} =\pm \hat{{\bf x}},\pm \hat{{\bf
y}}$ unit vectors, where $\epsilon_d$ is the d-electron energy
level and $V_h$ is the hybridization of the moments with the
conduction electrons. As a modification to the coupling term in
the ordinary Anderson impurity model, we here introduce a
site-dependent phase factor $\phi_i$, which is found to be quite
crucial in determining  the energy level and even the existence of
the impurity state if  $V_I$ is not so strong. Note that under the
meanfield decoupling scheme of Ref.
\cite{ubbens-prb-46-8434-1992}, the present model for
$H_{\text{mag}}$ is essentially equivalent to the magnetic
impurity(Kondo) model in Ref.\cite{polkovnikov-prl-86-296-2000},
where it is written as $\sum_i K_i {\bf S}\cdot{\bf s}(i)$ with
the summation being over the NN sites of the impurity, and ${\bf
S}=\sum_{\alpha\beta}d_{\alpha}^{\dagger}{\bf
\sigma}_{\alpha\beta}d_{\beta}/2$ and ${\bf
s}=\sum_{\alpha\beta}c_{\alpha}^{\dagger}{\bf
\sigma}_{\alpha\beta}c_{\beta}/2$ representing respectively the
spin operators of the magnetic impurity and the conduction
electrons. This is because that this spin-exchange term can be
transformed by the mean-field decoupling scheme in Ref.
\cite{ubbens-prb-46-8434-1992}: $K\sum_i {\bf S}\cdot{\bf
s}(i)\rightarrow
-3K/8\sum_{\sigma}[\chi_{i}^{*}c_{i\sigma}^{\dagger}d_{\sigma}+\text{h.c.}]$,
where a {\it complex} Hubbard-Stratonovich field $\chi_i=\langle
c_{i\sigma}^{\dagger}d_{\sigma} \rangle$ is introduced as an
effective hopping amplitude between the d-level and conduction
electrons. At this stage,
 one can see clearly that the
phase factor $\phi_i$ introduced in our model and
that of the field $\chi_{i}$ are closely related. Assuming the
four-fold rotational symmetry with respect to the impurity site,
there are four inequivalent arrangements of
$\phi_{\boldmath{\mbox{$\tau$}}}$, that is
\begin{equation}
\phi_{\boldmath{\mbox{$\tau$}}}^{(m)}=m\times\theta({\boldmath{\mbox{$\tau$}}}),
\label{eq_phi}
\end{equation}
with $m=0, 1, 2, 3$, $\theta(\pm\hat{{\bf x}})=(\pi/2)\mp(\pi/2)$,
and $\theta(\pm\hat{{\bf y}})=\pm(\pi/2)$. In Ref.
\cite{polkovnikov-prl-86-296-2000}, a nontrivial $d$-wave pattern
of $\chi_{i}$ has been found to have the lowest free energy
saddle point, whose phase corresponds to $m=2$, matching the
underlying symmetry of local bond pairing potential
$\Delta_{0,i}$. When the vortex exists, the symmetry of the underlying
pairing potential would change from the local $d$-wave pattern to that
responding to the winding of phase. Therefore, it appears  reasonable to
expect
that $\phi$ might change from $m=2$ to $m=1$ as  a vortex
 is pinned at the impurity site. We find that such a response of $\phi$ to the magnetic field
is quite crucial in keeping the impurity state in the vortex core.

In the treatment of the extended Anderson impurity model at low
temperatures, $U_d$ is assumed to be infinite as usual, which forbids
double occupancy of electrons on the  d-level. Therefore, the
slave-boson mean-field theory
\cite{barnes-jphys-6-1375-1976,coleman-prb-29-3035-1984} can be
applied as in Refs.
\cite{zhu-prb-64-060501-2001,zhang-prl-86-704-2001} where the
d-electron operator is written as
$d_\sigma^{\dagger}=f_{\sigma}^{\dagger}b$ with $f_{\sigma}$ the
spin-carrying fermion operator and $b$ the holon operator.
Furthermore, the single occupancy constraint
$\sum_{\sigma}f_{\sigma}^{\dagger}f_{\sigma}+b^{\dagger}b=1$
should be obeyed. At the mean-field level, the holon operators $b$
and $b^{\dagger}$ are approximated by a $c$-number $b_0$ and the
constraint is enforced on average by introducing a Lagrange
multiplier $\lambda_0$. Accordingly, the mean-field
$H_{\text{imp}}$ becomes
\begin{equation}
H_{\text{imp}}=\sum_{\sigma}\widetilde{\epsilon_d}f_{\sigma}^{\dagger}f_{\sigma}+
\sum_{{\bf
r}_i={\boldmath{\mbox{$\tau$}}},\sigma}\widetilde{V_h}(i)(e^{\phi_i}
c_{i\sigma}^{\dagger}f_{\sigma}+\text{h.c.})+ V_I
c_{0\sigma}^{\dagger}c_{0\sigma}+\lambda_0(b_0^2-1)
\end{equation}
with renormalized parameters
$\widetilde{\epsilon_d}=\epsilon_d+\lambda_0$ and
$\widetilde{V_h}=V_h b_0$. By applying the self-consistent
mean-field approximation and performing the Bogoliubov
transformation, diagonalization of the Hamiltonian can be
achieved by solving the following
Bogoliubov-de Gennes(BdG) equations:
\begin{equation}
\sum_{j}\left(
\begin{array}{cc}
H_{i,j} & \Delta _{i,j} \\
\Delta _{i,j}^{\ast } & -H_{i,j}^{\ast}
\end{array}%
\right) \left(
\begin{array}{c}
u_{j}^{n} \\
v_{j}^{n}%
\end{array}%
\right) =E_{n}\left(
\begin{array}{c}
u_{j}^{n} \\
v_{j}^{n}
\end{array}%
\right)  \label{BdG1}
\end{equation}%
where $u^{n},v^{n}$ are the Bogoliubov quasiparticle amplitudes \
with corresponding eigenvalue $E_{n}$ and $H_{i,j}=-\delta({\bf
r}_i+{\boldmath{\mbox{$\tau$}}}-{\bf r}_j)[t_{i,j\neq i_d
}-(\delta_{i,i_d}+\delta_{j,i_d})\widetilde{V_h}e^{i\phi_{\boldmath{\tiny\mbox{$\tau$}}}}]-
\delta_{i\neq i_d,j}[\mu-\delta({\bf r}_i-{\bf
r}_0)V_I]+\delta_{i,i_d}\widetilde{\epsilon_d}$.
Here,  $i,j$ represent the index of the 2D lattice sites
with $i_d$ as the  index for the magnetic moment residing
on the NN sites of the doped impurity. $\Delta_{i,j}$ is defined between a pair of NN sites
on the 2D
lattice and is calculated according to the self-consistent condition:
\begin{equation}
\Delta_{i,j}=\frac{V_d}{2}\sum_{n}(u_i^n v_j^{n*}+u_j^n
v_i^{n*})\tanh(\frac{E_n}{2k_B T})
\end{equation}
and
\begin{equation}
b_0^2=1-2\sum_n\{|u^n_{i_d}|^2f(E_n)+|v^n_{i_d}|^2[1-f(E_n)]\}
\label{eq_b0}
\end{equation}
Once the BdG Eq. \ref{BdG1} is solved  self-consistently, the
quasiparticle spectrum can be obtained and the LDOS proportional
to the differential tunneling conductance observed in STM
experiments is given by
\begin{equation}
\rho({\bf r}_i,E)=-\sum_{n}[|u_i^n|^2 f^{\prime}(E_n-E)+|v_i^n|^2
f^{\prime}(E_n+E)]
\end{equation}
where $f(E)$ represents
the Fermi distribution function.

{\it Unitary Impurity Model.}--- First, we study the variation of
the impurity state when a vortex line is sitting on the impurity
site within a unitary impurity model. Exact diagonalization is
applied and the BdG equations are solved self-consistently. In
studying the electronic structure of vortex lattice in $d$-wave
superconductors, a magnetic unit cell which accommodates two vortices
is usually employed in numerical studies
\cite{wang-prb-52-r3876-1995,takigawa-prl-83-3057-1999,zhu-prl-87-147002-2001,wang-prl-87-167004-2001,han-prb-65-064527-2002}.
 Such a method has been applied to treat impurity
effect in the mixed state of $s$-wave
\cite{han-prb-62-5936-2000}and $d$-wave
\cite{zhu-cond-mat-0109503-2001} superconductors. The primitive
translation vectors of the magnetic unit cell are ${\bf R}_x=aN_x
\hat{x}$ and ${\bf R}_y=aN_y \hat{y}$, where $a$ is the lattice
constant and will be set as unity. The pairing potential winds by
$4\pi$ around the magnetic unit cell. The quasiparticle
amplitudes $u^n$ and $v^n$ are classified by the magnetic Bloch
quasi-momentum ${\bf k}$.

The parameters we choose are $\mu=-0.2 t$, $V_d=2.2 t$ which give
rise to $\Delta_d=0.274 t$ (accordingly
$\Delta_{\text{max}}\simeq t$) and $T_c=0.45 t$.
We here intentionally take a relatively large amplitude of the energy gap $\Delta_{\text{max}}$
 with respect to real materials,
because a large energy gap can sufficiently lower the LDOS peak
corresponding to the vortex core states and in our case a broad
structure is achieved near the vortex center \cite{comment1},
which enables us to distinguish the LDOS peak of the unitary
impurity from that corresponding to the vortex core states. The
on-site attractive scattering potential is chosen as $V_I=-10 t $.
In Fig. \ref{fig1} the LDOS at the NN site of Zn site is plotted.
In the absence of  magnetic field(dashed line), a sharp peak is
found at $E_0/\Delta_{max}=-0.02$ corresponding to the impurity
resonance. In the inset, the spatial distribution of the LDOS at
resonance energy, $\rho({\bf r},E_0)$ is shown, indicating that
the spectral weight is concentrated at the NN sites of the
impurity while vanishingly small at the impurity site. These
results are qualitatively consistent with previous numerical
investigations. When a vortex  is pinned right on the impurity
site, from Fig. \ref{fig1}(solid line)  we can see that the  LDOS
peak at the NN site of Zn is significanly suppressed and mixes
with the vortex core states without identification of impurity
resonance state any more. As a strong scattering potential, the
unitary impurity can drive the spectral weight on it mainly to its
NN sites no matter vortex is present or not. However, in the
absence of vortex, the LDOS at a specific energy is raised,
leading to a resonance state; while in the presence of vortex, the
LDOS on NN sites is increased in a wide energy range as seen in
Fig. \ref{fig1} and thus it is more reasonable to state that the
vortex core states are strengthened by the impurity scattering.
Broadening of the impurity resonance state by the magnetic field
exists when the supercurrent of the vortex disturbs the energy
spectrum of $d$-wave superconductors {\it via} Doppler shift
\cite{samokhin-prb-64-024507-2001}. However, in our case the
smearing of the impurity resonance state is not induced by the
supercurrent because the impurity is on the vortex center where
supercurrent is zero. We attribute it to the nature of topological
singularity of the vortex center where the phase of pairing
potential varies arbitrarily, which makes the vortex center itself
a strong scatter of electrons \cite{nielsen-prb-51-7679-1995}.
Therefore, it is the vortex that modifies the background
electronic structure which is responsible for the impurity
resonance and accordingly makes the impurity resonance
indiscernible.

{\it Extended Anderson Impurity Model.}---  We now address the
impurity state in the presence of Kondo screening of the magnetic
moments described by the extended Anderson impurity model. First
we consider the case in the absence of applied magnetic field. We
set $m=2$ in Eq. (\ref{eq_phi}) in consideration of  free energy
minimization \cite{polkovnikov-prl-86-296-2000}. The exact
diagonalization is performed on a $24\times 24$ square lattice,
which is seen to be sufficiently large due to the localized
feature of the impurity state. The scattering potential $V_I$ is
simplified as zero (The role of nonzero $V_I$ will be discussed
later). The bare d-level is set as $\epsilon_d=-4t$ and the
coupling strength $V_h=2t$(Kondo screening regime). The value of
the Lagrange multiplier $\lambda_0$ is determined by minimizing
the free energy of the system as indicated in the inset of Fig.
\ref{fig2} and the corresponding value of $b_0$ is calculated
self-consistently according to  Eq. (\ref{eq_b0}). In the inset of
Fig. \ref{fig2},  $\Delta F$ as a function of
$\lambda_0$ is plotted, where $\Delta F$ represents the change of
the free energy when the local magnetic moments couple with
conduction electrons. We find that for the chosen parameters the
lowest free energy occurs when $\lambda_0\simeq 5.6 t$ with
$b_0\simeq 0.64$. Figure \ref{fig2} gives LDOS's on the Zn site,
its NN site and NNN site. There is a sharp peak at the energy
$E_0/\Delta_{max}=-0.04$, which is well consistent with STM
measurement. Right on the Zn site, the peak is strongest and
local maximum peaks are on the NNN sites; while on the NN sites
peaks are rather weak and at $-E_0/\Delta_{max}=0.04$. The spatial
distribution of LDOS at $E_0$ is shown in the inset of Fig.
\ref{fig2}, also in agreement with STM spectra
\cite{pan-nature-403-746-2000} and a theoretical study based on
another approach on an essentially equivalent model
\cite{polkovnikov-prl-86-296-2000}. Our analysis shows that  the
impurity state induced by the magnetic moment is a bound state
with a very short attenuation length $\sim \sqrt{2}a$.

Recent STM measurements \cite{yeh-prl-87-087003-2001} of impurity
states induced by Zn$^{++}$ and Mg$^{++}$ doped in
YBa$_2$CuO$_{6.9}$ reported that the resonance energies are at
$(-10\pm 2)$ and $(4\pm 2)$ meV, being not necessarily near
zero bias. Based on our model, with the fixed coupling strength
e.g. $V_h=2t$, both the bare d-level $\epsilon_d$ and
on-site scattering potential $V_I$ could adjust the resonance position of the
impurity state. When $V_I=0$ and $\epsilon_d$ ranges from $0$ to
$-4 t$, $E_0/\Delta_{max}$ varies from $-0.18$ to $-0.04$.
An attractive(negative) $V_I$ drives the energy position below the
Fermi level further (e.g. when $V_I=-1t$, $E_0/\Delta_{max}$
varies from $-0.42$ to $-0.26$ as $\epsilon_d$ ranges from $0$ to
$-4 t$), while repulsive one makes the resonance energy
approach to or even across the Fermi level and become
positive (e.g. when $V_I=t$, $E_0/\Delta_{max}=0.25$ with
$\epsilon_d=-4 t$). However, too large $V_I$ ($\gg t$) can
affect the spin-induced impurity state strongly and double resonance peaks
\cite{zhu-prb-63-020506-2000} are found in the
unitary limit, which disagrees with the STM
spectra.

When a vortex is pinned on the impurity site, we expect that
there is a rearrangement of
$\phi_{\boldmath{\mbox{$\tau$}}}^{(m)}$ according to the winding
of the phase of the order parameter so that the phases of the
Hubbard-Stratonovich field $\chi_i$ and the local bond pairing
potential between NN sites $\Delta_{0,i}$ have the best match.
Therefore, $m$ will vary from $2$ to $1$ when a
 vortex is present. This is quite important as  other
choices of $m$ ($0,2,3$)  are found to result in the disappearing of the
impurity state. The numerical results are shown in Fig. \ref{fig3}, which plots the
LDOS as a function of energy on the impurity site.  A
strong single peak is seen near the Fermi level, indicating that
 the impurity state still survives under the field. Being different from the
case without vortex, the peak is lowered slightly and moved to
positive bias $E_0/\Delta_{max}=0.02$, while  the spatial pattern of the impurity
state is not modified(see the inset of Fig.
\ref{fig3} ).

In summary, we have compared the extended Anderson impurity model
with the unitary impurity model by studying numerically the
impurity state of a $d$-wave superconductor as a vortex is pinned on
the impurity site. We find that the impurity resonance governed by the
scattering potential mechanism is  sensitive to the presence
of vortex, while the impurity bound state generated from
the magnetic moments mechanism seems rather robust
to the field.
Therefore, it may be helpful to clarify the dominant mechanism of
the impurity state in Zn or Mg impurities doped cupates
by examining the existence of
the impurity resonance, the variation of its energy, and even
the STM spatial pattern  in the presence of  a strong magnetic field.


We thank Profs. C. S. Ting, J. L. Zhang, and Dr.  Y. Chen   for helpful discussions. The work was
supported by a RGC grant of Hong Kong under Grant No. HKU7144/99P
and the 973-project of the Ministry of Science and Technology of
China under Grant Nos. G1999064602\&G1999064603. ZDW acknowledges
partial support from the Texas Center for Superconductivity at the
University of Houston.

\begin{figure}
\caption{LDOS versus energy at
an NN site within the unitary impurity model for the cases (i)
without vortex(dashed line), (ii) a vortex  pinned at the impurity
site (solid line),
 and (3) a vortex core without impurity (dotted line). The bulk density of states
is also plotted for reference (dash-dotted line). Inset shows the
spatial distribution of the LDOS at energy
$E_0/\Delta_{max}=-0.02$ in a $16\times 16$ region for case (i). }
\label{fig1}
\end{figure}

\begin{figure}
\caption{LDOS versus energy  at the impurity site (solid
line), NNN site (dashed line), and NN site (dotted
line) under zero field within the extended Anderson impurity
model. The bulk density of states is also plotted for reference
(dash-dotted line). Inset (a) shows the variation of $\Delta
F$ as a function of $\lambda_0$ and (b) gives the spatial
distribution of the LDOS at energy $E_0/\Delta_{max}=-0.04$. }
\label{fig2}
\end{figure}

\begin{figure}
\caption{Same as Fig. \ref{fig2} except that a  vortex is
pinned on the impurity site. Inset (b) gives the spatial
distribution of the LDOS at energy $E_0/\Delta_{max}=0.02$. }
\label{fig3}
\end{figure}

\newpage
\thispagestyle{empty}
\centerline{\epsfig{figure=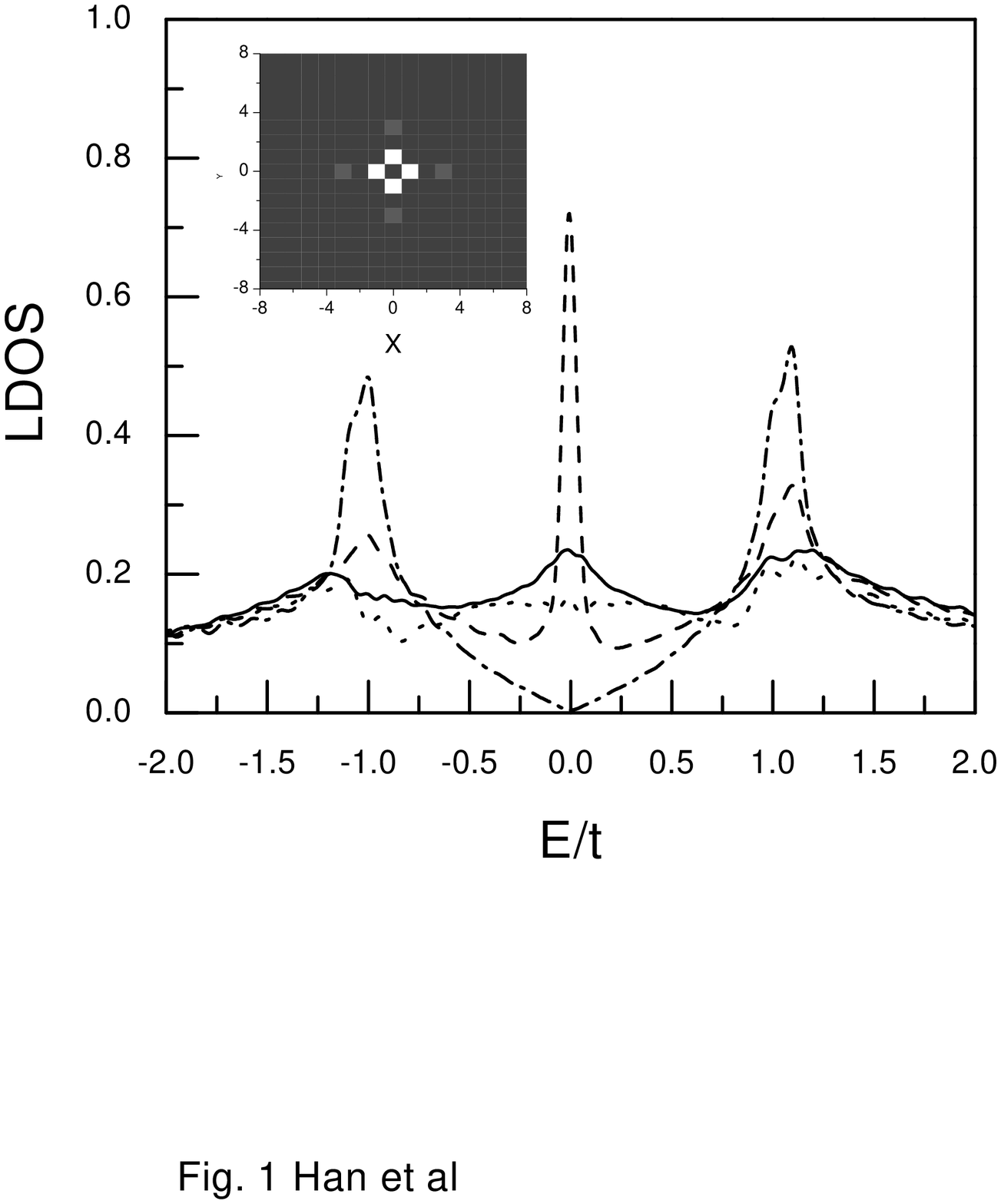}} %
\newpage
\thispagestyle{empty}
\centerline{\epsfig{figure=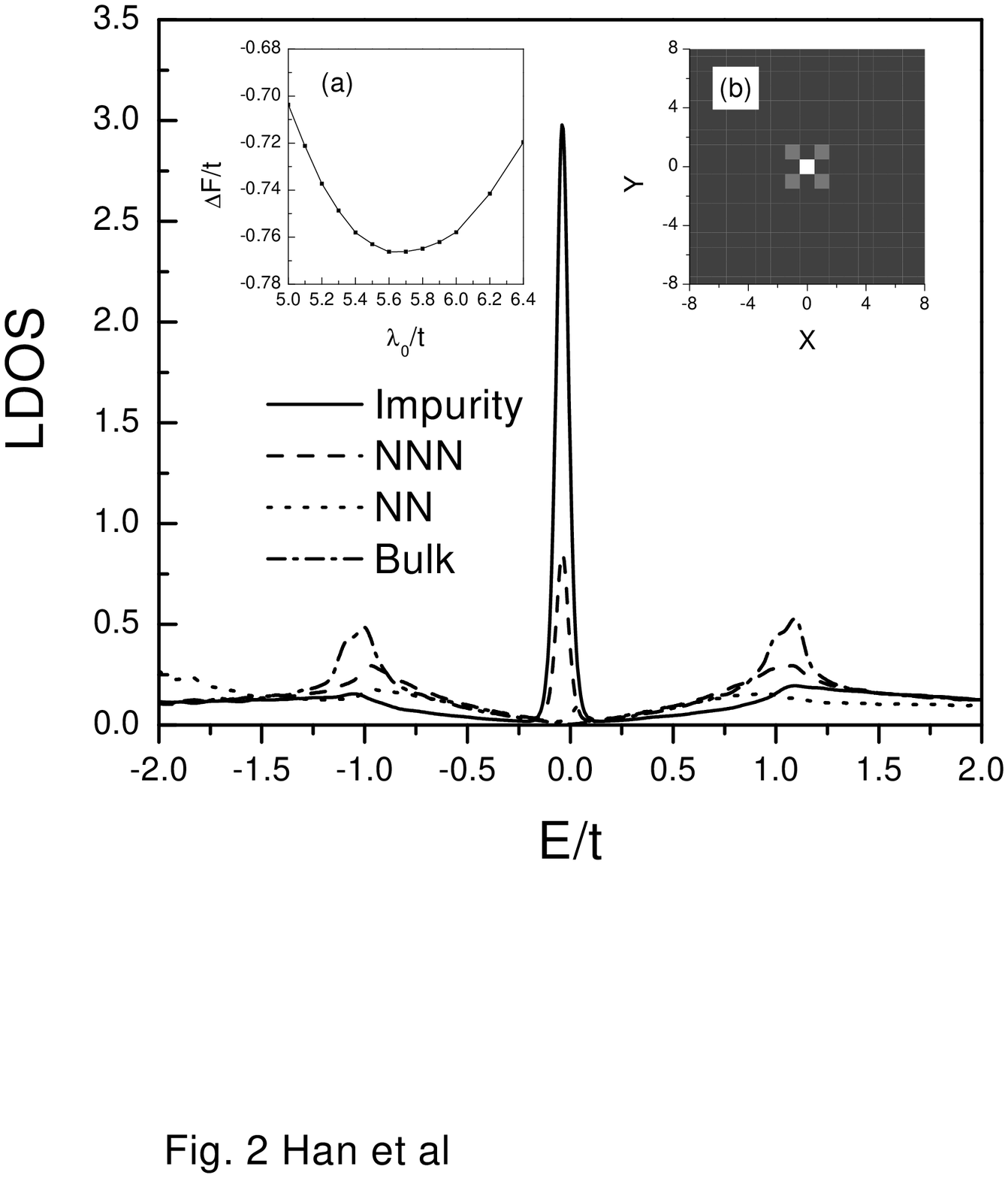}} %
\newpage
\thispagestyle{empty}
\centerline{\epsfig{figure=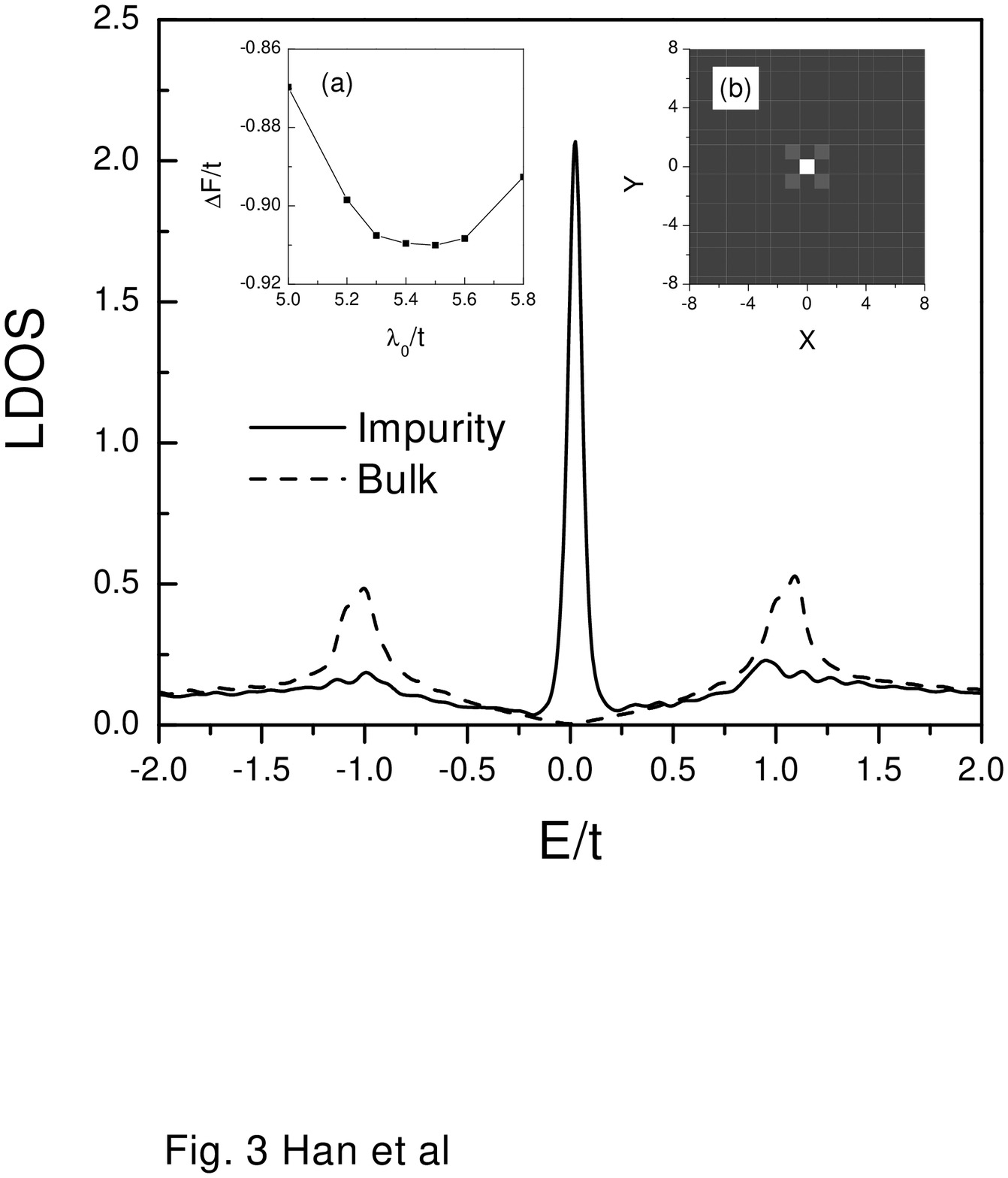}}
\thispagestyle{empty}

\end{document}